\documentclass[12pt]{iopart}
\usepackage{bm}
\usepackage[T1]{fontenc}
\usepackage[latin1]{inputenc}
\usepackage{graphicx}
\usepackage{amssymb}
\usepackage{esint}

\begin{document}

\title{Scattering in one-dimensional heterostructures described by the Dirac 
equation}
\author{N. M. R. Peres}

\address{Center of Physics  and  Department of Physics,
             University of Minho, P-4710-057, Braga, Portugal}

\date{\today}
\begin{abstract}
We consider electronic transport accross one-dimensional heterostructures described by the
Dirac equation. We discuss the cases where both the velocity and the mass are
position dependent. We show how to generalize the Dirac Hamiltonian
in order to obtain a Hermitian problem for spatial dependent velocity.
We solve exactly the case where the position dependence of both
velocity and mass is linear. 
In the case of velocity profiles, it is shown that there is
no backscattering of Dirac electrons. 
In the case of the mass profile backscattering exists. In this case,
it is shown that the linear mass profile induces less backscattering than the abrupt step-like 
profile. Our results are a first step to the study of similar problems in graphene.
\end{abstract}



\section{Introduction}

Most of the accumulated knowledge about the physics of heterostructures 
assumes that the electrons in these materials are effectively described by the
Shr\"odinger equation, with a position dependent mass. The common belief is that 
Dirac equation is of no use in condensed matter physics (spin-orbit coupling can be treated using the
Pauli version of the Schr\"odinger equation). The discovery of graphene \cite{novo1,pnas}
changed drastically this perspective. Condensed matter physicists are now facing a condensed matter
system where the effective low-energy model for the quasiparticles is that of a ultrarelativistic, i. e. 
massless, 
Dirac equation, albeit with an effective Fermi velocity that is much lower than the velocity
of light. In graphene the Fermi velocity is $v_F=c/300$, with $c$ the velocity of light. 
In fact, since the isolation of graphene crystallites \cite{novo1,pnas,rmp} that a renovated
interest in the properties of the massless Dirac equation in 2+1 dimensions started to
emerge \cite{rmp,scientometrics}. Although our interest in this paper is on the backscattering
properties of electrons effectively  described by the Dirac equation in 1+1 dimensions, as an effective
low-energy theory of the electronic properties of a quasi-one-dimensional solid, we shall revise some of the properties of Dirac electrons in the context of graphene, given the accumulated amount of evidence
for Dirac fermions in graphene. 

How can the low-energy description of a solid be given by an effective Dirac-like equation?
In one-dimensional physics, this takes place due to the linearization of the spectrum close to
the Fermi momentum, which introduces right- and left-movers with a linear energy-momentum relation.
Another possibility, that we discuss in the bulk of the paper, is by having two atoms per unit cell
with special type of hopping. In two-dimensions, this possibility has been realized by graphene. 
The
electrons in graphene are 
confined to move within the $\pi-$orbitals of system, formed
from the overlap of the $2p_z$ atomic orbitals of a single carbon atom. The transport and
 optical properties of graphene are determined essentially by the behavior of the
$\pi-$electrons. The electronic density is such that graphene has one $\pi-$electron
per carbon atom, and is therefore a half filled system. 
It turns out that the band structure of graphene around the {\it Fermi surface}
is highly non-standard, since the relation between the momentum and the energy is linear,
with a proportionality  coefficient given by $v_F\simeq 1\times 10^6$ m/s and termed
Fermi velocity. In fact, the {\it Fermi surface} is reduced to two points (the so
called Dirac points $\bm K$ and $\bm K'$)
in the
Brillouin zone, with a zero density of states at the Fermi-points.
This fact gives graphene its non-conventional properties.

Starting from a $\bm k\cdot\bm p$ approach \cite{davies},
adapted to degenerate bands off the $\Gamma-$point \cite{mele}, we obtain an
effective Hamiltonian  in real space, valid around the $\bm K$ point, having  
the form

\begin{equation}
 H = v_{F} {\bm \sigma} \cdot {\bm p}\,\,,
\label{HD2d}
\end{equation}
where ${\bm \sigma}=(\sigma_x,\sigma_y)$ is a vector using the Pauli matrices 
and ${\bm p}= -i\hbar (\partial_x,\partial_y)$.
The Hamiltonian (\ref{HD2d}) is nothing but the Dirac Hamiltonian for massless
particles in 2+1 dimensions, with an effective light-velocity $v_F$. Clearly, one expects that the physical properties of such
a system will be different from those where the Schr\"odinger equation is valid.
Note that this situation is an example of complex emergent behavior \cite{pwanderson}, since the original problem
was that of independent electrons (and therefore Schr\"odinger like) in a periodic
potential. 

Those electrons in graphene lying close to the Fermi points ($\bm K$ and $\bm K'$),
will have a different response to external potentials than those in other materials
described by the Schr\"odinger equation. In fact, for Schr\"odinger electrons, 
if we consider that the potential
varies at a given point in space from zero to a finite value $V_0$ 
(the so called step potential) there is always a finite fraction of impinging electrons
that are reflected back. More over, if $V_0$ is greater than the energy of the impinging
particles, there will be an exponential attenuated wave function in the region
where the potential is finite.

For Dirac electrons the situation is different. It was shown by Klein \cite{calogeracos}
that Dirac particles can not be confined by an arbitrary large potential $V_0$.
In fact, they
pass through strong repulsive potential without the exponential decay that
characterizes  Schr\"odinger particles. This is called Klein tunneling.
If the particles are massless and moving in one-dimension
the situation is even more dramatic, since
there is no backscattered probability flux, no matter how large the 
potential is \cite{NaturePhys,rmpBeenakker}.

In addition to external potentials, a particle propagating in solid state systems can 
face a situation where its effective mass changes in space. This possibility takes place
in heterostructures, where the mass of the particle is written as $m(\bm r)$, $\bm r$
being the position of the particle. In this situation, the Sch\"rodinger equation has to be 
modified, in other to comply with hermiticity and flux conservation. 

If the particles are described by the Dirac equation, the information on the material
properties is encoded both in the Fermi velocity \cite{mele} and in the mass. 
We can then imagine the possibility
of producing a herestructure where the Fermi velocity, $v_F(\bm r)$, and the mass change with the
position of the particle. As we show bellow,  even in this case, there will be not
backscattering if the particle is massless and the velocity is allowed to vary from point to point.
In the context of graphene physics, a position-dependent Fermi velocity was already considered
in the case of curved graphene \cite{maria}. Another possibility of producing 
both a Fermi velocity and a mass term that are position dependent is to subject the system to
strain. It has already been shown using {\it ab-inito} calculations, in the context of graphene physics, that strain leads to
gap formation \cite{strain1,strain2}. It is to be expected that for general strain, different
parts of the material presents local values of Fermi velocity and energy gap. Also in the
case of graphene, it was shown that depositing the material on top of Boron Nitride leads
to gap-formation \cite{Giovannetti07}.
 
In this paper we discuss several possibilities for the scattering of Dirac particles
through a region where both the velocity and the mass are position dependent. Although our motivation
is to study graphene strips \cite{zecarlos} in a latter publication, 
we start by giving two exact solutions for one-dimensional systems.
The generalization to a quasi-one-dimensional system, such as a narrow nano-wire, is easy to
obtained and will be given in a follow up publication. In order to see the differences
between the Schr\"odinger and the Dirac problems, we start by revising the case of 
one-dimensional heterosctructures described by the Schr\"odinger equations
and latter move the study of the Dirac case.

\section{Warming up: the Schr\"odinger electrons in 1D}

In order to compare how the scattering of Schr\"odinger electrons differs from that
of Dirac particles, we give a brief account of the scattering of electrons
by the interface of a heterostructure.
 
Let us consider the case of 1D Schr\"odinger electrons with position dependent 
effective mass $m(x)$,
such that $m(x)=m^-$, for $x<0$ and $m(x)=m^+$, for $x>0$.
The boundary conditions in this case are (see final paragraph of this section; for more general boundary conditions see Ref. \cite{harrison} )
\begin{eqnarray}
 \psi^-(0)&=&\psi^+(0)\,,\\
\frac{1}{m^-}\frac{d}{d\,x}\psi^-(0)&=&\frac{1}{m^+}\frac{d}{d\,x}\psi^+(0)\,.
\label{bcII}
\end{eqnarray}
We consider that the particles are moving from the left to the right. Therefore
the wave function is 

\begin{eqnarray}
\psi^-(x)&=&e^{ikx}+re^{-ikx}\,,\\
\psi^+(x)&=& te^{ipx}\,, 
\end{eqnarray}
where the values of $k$ and $p$ are fixed by energy conservation:
\begin{equation}
 E=\frac{\hbar^2 k^2}{2m^-}=\frac{\hbar^2 p^2}{2m^+}\,.
\end{equation}
The reflectance  $\vert r\vert ^2$ is determined from the boundary 
conditions as

\begin{equation}
\vert r\vert^2 =\frac {(m^+k-m^-p)^2} {(m^+k+m^-p)^2}\,,
\end{equation}
which has the limiting value $\vert r\vert^2\rightarrow 1$ for $m^+\rightarrow \infty$.
The transmittance coefficient  $\vert t\vert ^2$  is 

\begin{equation}
 \vert t \vert^2 = \frac{4(m^+k)^2}{(m^+k+m^-p)^2}\,,
\end{equation}
We can check now that the boundary conditions used for the wave function are the
correct ones, since they satisfy the flux current conservation.
Using absolute values, we have indeed
\begin{equation}
 j^{inc} = j^{reft} + j^{trans}\,,
\end{equation}
with  $j^{inc}=\hbar k / m^+$, $j^{reft}=\vert r\vert^2\hbar k / m^+$, and 
$j^{trans}=\vert t\vert^2\hbar p / m^-$.

The problem just described represents an heterostructure, where two materials
are {\it glued} together (due to similar lattice constants), 
having different effective masses in the two materials.
The used boundary conditions are obtained writing the Hamiltonian
as
\begin{equation}
  H = -\frac{\hbar^2}{2m}\frac{d\,}{d\,x}\frac{1}{m(x)} \frac{d\,}{d\,x}\,.
\label{HS}
\end{equation}
This choice defines an Hermitian problem
\footnote{The Eq. (\ref{HS}) is a particular case of a Sturm-Liouville operator, which has
the general form:
$$-\frac {d}{d\,x}\left(
p(x)\frac {d}{d\,x}
\right) + q(x)y=\lambda w(x)y\,,$$ 
where $\lambda$ is to be determined from the boundary conditions,
and $p(x)$, $q(x)$, and $w(x)$ are given functions.
Sturm-Liouville problems are Hermitian.}
 since $\int dx (H\psi)^\ast\psi=\int dx \psi^\ast H\psi$. The construction
of the (\ref{HS}) guaranties that the total probability flux is conserved.
The boundary condition (\ref{bcII}), follows immediately from the writting the
Hamiltonian in the Sturm-Liouville form.

\section{Dirac electrons in 1D}

Let us now study the case of the Dirac Hamiltonian. We are interested in cases where both the
velocity and the mass of the particles depend on their position in space. We study first the case of massless Dirac particles,
like those present on graphene. Latter we add up the mass term.
The Hamiltonian for a massless quasiparticle  described by an effective Dirac equation, with 
an effective velocity of light $v_F$, is given by

\begin{equation}
 H=v_F\sigma_x\frac{\hbar}{i}\frac{d}{d\,x}\,,
\label{HD}
\end{equation}
where $\sigma_x$ is the $x$ Pauli matrix. 
Let us now consider that we make an
heterostructure, made of two different materials, both of them described by an effective
Dirac equation, such as (\ref{HD}). An example could be a strip of the material where part of it
is subjected to strain and other part is strain free; this leads to different velocities
in the two parts of the material.
This situation calls for a model where $v_F$
is position dependent: $v_F=v_F(x)$. Although this situation does not make sense 
in high-energy physics, it is quite conceivable in condensed matter physics, since
the value of $v_F$ is determined by the material under consideration -- different materials
can have different Fermi velocities. 

The trivial replacement $v_F\rightarrow v_F(x)$ renders the problem
non-Hermitian, as can be seen by applying the definition given above, when we discussed
the Schr\"odinger case. 
There is however a way out. It is a simple matter to show that the operator

\begin{equation}
 H=\sqrt{v_F(x)}\sigma_x\frac{\hbar}{i}\frac{d}{d\,x}\sqrt{v_F(x)}\,,
\label{HDII}
\end{equation}
is Hermitian and reduces to Eq. (\ref{HD}) in the particular case $v_F(x)=v_F$.
Note that the derivative will act on the product $\sqrt{v_F(x)}\psi(x)$. 
We stress that we are not studying the scattering of relativistic particles in condensed matter
systems, but instead describing the scattering of particles represented by an effective low
energy model that is formally equivalent to the Dirac equation (a situation that takes place in graphene).
It would be interesting
to derive Eq. (\ref{HDII}) from a microscopic Hamiltonian.
In \ref{Ap00} we give a tight-binding model whose effective low-energy theory
is given by Eq. (\ref{HDII}).
The  problem (\ref{HDII}) has spinorial wave function of the form $\psi^\dag=(\psi_1^\ast, \psi_2^\ast)$ and the
probability flux is computed as  $ S_x=v_F(x)\psi^\dag \sigma_x\psi$,
as can be shown using the traditional derivation of computing the time change of the
probability density \cite{shiff}.

The eigenproblem $H\psi=E\psi$, with $H$ given by Eq. (\ref{HDII}),
corresponds to two coupled first-order differential equations of the
form
\begin{eqnarray}
\label{eq:firstI}
\sqrt{v(x)}\frac{\hbar}{i}\frac{d\,[\sqrt{v(x)}\psi_2(x)]}{d\,x}=E\psi_1(x)\,,\\ 
\label{eq:firstII}
\sqrt{v(x)}\frac{\hbar}{i}\frac{d\,[\sqrt{v(x)}\psi_1(x)]}{d\,x}=E\psi_2(x)\,. 
\end{eqnarray}
It is straightforward  to show that the two first-order differential equations
given above can be put in Sturm-Liouville form:
\begin{equation}
 -\frac{d}{d\,x}[v_F(x)\frac{d}{d\,x}y_1(x)]=\frac{E^2}{v_F(x) \hbar^2}y_1(x)\,,
\label{SL}
\end{equation}
with $y_1(x)= \sqrt{v_F(x)}\psi_1(x)$, $p(x)=v_F(x)$, $\lambda=E^2/\hbar^2$, and $w(x)=1/v_F(x)$,
and $\psi_2(x)$ obtained from
\begin{equation}
 \psi_2(x) = \frac{\hbar}{iE}\sqrt{v_F(x)}\frac{d}{d\,x}[\sqrt{v_F(x)}\psi_1(x)]\,.
\label{psi2}
\end{equation}
 If the velocity profile changes continuously, one expects that
the continuity of the wave function should be valid. If the velocity profile
changes abruptly at a given point in space, say at $x=0$, from $v(x)=v_-$, for $x<0$,
to $v(x)=v_+$, for $x>0$,
 one has to use Eqs. (\ref{eq:firstI})
and (\ref{eq:firstII}) to derive the boundary conditions.
Defining $y_2(x)= \sqrt{v_F(x)}\psi_2(x)$ Eqs. (\ref{eq:firstI})
and (\ref{eq:firstII})  can be written as
\begin{eqnarray}
\label{eq:firstIb}
\frac{\hbar}{i}\frac{d\,[y_2(x)]}{d\,x}=E\frac{y_1(x)}{v(x)}\,,\\ 
\label{eq:firstIIb}
\frac{\hbar}{i}\frac{d\,[y_1(x)]}{d\,x}=E\frac{y_2(x)}{v(x)}\,. 
\end{eqnarray}
Integrating Eqs. (\ref{eq:firstIb})
and (\ref{eq:firstIIb}) using a symmetric infinitesimal interval around $x=0$
one obtains the condition
\begin{equation}
y_i(0^-)=y_i(0^+)\,, 
\label{bc}
\end{equation}
with $i=1,2$. Note that condition (\ref{bc}) implies the discontinuity of the wave function.
Let us now move to the solution of several particular cases.
\subsection{The step-like velocity profile}
\label{step-like}

Let us consider the case 
\begin{equation}
 v_F(x)=v^-_F\,\theta(-x)+v^+_F\,\theta(x)\,.
\label{Vprofile}
\end{equation}
The solution of the Dirac equation reads
\begin{equation}
 \psi_-(x)
=\left(
\begin{array}{c}
1\\
1
\end{array}
\right)e^{iqx}
+
r
\left(
\begin{array}{c}
1\\
-1
\end{array}
\right)e^{-iqx}\,,
\label{psiinc}
\end{equation}
and
\begin{equation}
 \psi_+(x)
=
t\left(
\begin{array}{c}
1\\
1
\end{array}
\right)e^{ipx}\,,
\label{psitrans}
\end{equation}
with $E=v_F^-q\hbar $ and $E=v_F^+p\hbar$. 
The above wave functions were obtained by solving the Dirac equation
for $x\gtrless 0$, where the velocity is constant.
Using these solutions and the boundary condition (\ref{bc}) one obtains
\begin{equation}
 \left\{
\begin{array}{c}
 \sqrt{v_-}(1+r)=\sqrt{v_+}t\,,\\
\sqrt{v_-}(1-r)=\sqrt{v_+}t\,,
\end{array}
\right.
\label{coeffs}
\end{equation}
which is satisfied only for $r=0$.
Had we consider the general case of both a velocity profile (\ref{Vprofile}) and a potential
profile of the form 
\begin{equation}
 V(x)=V_0\,\theta(x)\,,
\end{equation}
under the condition $E>V_0$,
the boundary conditions would still give $r=0$. This result is called Klein 
tunnelling \cite{calogeracos}.
If we consider the same velocity and potential profiles, but now take the energy
$E<V_0$, the boundary conditions still gives the result  (\ref{coeffs}),
but the wave function of the propagating mode for $x>0$ is now different and given by
 \begin{equation}
\psi_+(x)=t
\left(
\begin{array}{c}
 1\\
1
\end{array}
\right)e^{-ipx}\,.
\end{equation}
In this case the result will also be $r=0$. The conclusion is that it is not
possible to backscatterer massless Dirac electrons 
with a step-like velocity and potential profiles in 1D.

\subsection{The linear velocity profile with massless particles}
\label{linear}

We have seen that an abrupt change of the velocity at the interface produces no 
reflected particles. Let us now study the case of a smooth change in
the velocity from $v_F^-$ to $v_F^+$. (The result can be guessed from the outset!)
To that end, we choose the profile

\begin{equation}
v_F(x)=v^-_F\,\theta(-\delta-x)+(\bar v+x\Delta )\,\theta(\delta-\vert x \vert)+
v^+_F\,\theta(x-\delta)\,,
\end{equation}
where we have defined $\bar v$ and $\Delta$ as 
\begin{eqnarray}
 \bar v &=\frac{v^-_F+v^+_F}{2}\,,\\
\Delta &= \frac{v^+_F-v^-_F}{2\delta}\,.
\end{eqnarray}
For the cases $x<-\delta$ (region $I$)
or $x>\delta$ (region $III$)
the Fermi velocity is constant and the solution of the Dirac
equation  is elementary, as we have seen before. The interesting case is therefore
the region $\vert x\vert < \delta$ (region $II$). In this case we have to use 
the Dirac equation in the form (\ref{HDII}). Explicitly we have to solve the problem
$H\psi=E\psi$ with $H$ given by

\begin{equation}
 H=\sqrt{\bar v + x\Delta }\,\sigma_x\frac{\hbar}{i}\frac{d}{d\,x}\sqrt{\bar v + x\Delta }\,.
\label{HDexample}
\end{equation}
Writting the differential equations satisfied by the spinors, we obtain for the $\psi_1$ spinor
(considering the substitution $y=\sqrt{\bar v + x\Delta }\psi_1$) the equation
\begin{equation}
 -v_F^2(x)\frac{d^2\,y}{d\,x^2}-\Delta v_F(x)\frac{d\,y}{dx}=\epsilon^2 y\,,
\end{equation}
with $\epsilon^2=E^2/\hbar^2$. Making the replacement
\begin{equation}
 z = \frac{\bar v +x \Delta}{\bar v}\equiv \theta(x)\,,
\end{equation}
we obtain the result
\begin{equation}
 z^2\frac{d^2\,y}{d\,z^2}+z\frac{d\,y}{dz}+\nu^2 y=0\,,
\label{HDyz}
\end{equation}
with $\nu^2=\epsilon^2/\Delta^2$. Making the additional replacement
$\omega=\ln z$, Eq. (\ref{HDyz}) is reduced to that of the harmonic oscillator
\footnote{One should note that the solution of Eq. (\ref{HDyz}) can also be obtained
by noticing that $y=z^m$ is a solution if $m^2=-\nu^2$, leading to
$y(z)=Az^{i\nu}+Bz^{-i\nu}$. }
\begin{equation}
 \frac{d^2\,y}{d\,\omega^2}+\nu^2 y=0\,.
\label{HDomega}
\end{equation}
The general solution of Eq. (\ref{HDomega}) is elementary and from it one obtains $y(x)$ given by
\begin{equation}
 y(x)=A\sin[\nu L\theta(x)] + B\cos[\nu L\theta(x)],
\end{equation}
with 
\begin{equation}
 L\theta(x) = \ln \frac{\bar v +x \Delta}{\bar v}\,.
\end{equation}
The component $\psi_1$ of the spinor is obtained from $\psi_1(x)=y(x)/\sqrt{v_F(x)}$, and
 $\psi_2$ is obtained using Eq. (\ref{psi2}) and reads
\begin{equation}
 \psi_2(x)=\frac{\hbar}{iE}\sqrt{v_F(x)}\frac {d\,y(x)}{d\,x}\,.
\end{equation}
The boundary conditions are $\psi^I(-\delta)=\psi^{II}(-\delta)$ and 
$\psi^{II}(\delta)=\psi^{III}(\delta)$, and can be written as
\footnote{Note that the dimensions of $A$ and $B$ are $\sqrt{L/T}$, and
$\lambda^\pm$ has dimensions of $\sqrt{T/L}$.}

\begin{equation}
 \left(
\begin{array}{cccc}
-e^{i k\delta } & S_-/\sqrt{v^-_F}  &  C_-/\sqrt{v^-_F} & 0 \\
e^{i k\delta } & C_-\lambda^-  &  -S_-\lambda^+ & 0 \\
 0 & S_+/\sqrt{v^+_F}  &  C_+/\sqrt{v^+_F} & -e^{iq\delta} \\
 0 & C_+\lambda^-  &  -S_+\lambda^+ & -e^{iq\delta} 
\end{array}
\right)
\left(
\begin{array}{c}
r\\
A\\
B\\
t
\end{array}
\right)
=
\left(
\begin{array}{c}
e^{-i k\delta }\\
e^{-i k\delta }\\
0\\
0
\end{array}
\right)
\end{equation}
 with 
\begin{eqnarray}
 S_\pm&=&\sin[\nu L\theta(\pm\delta)]\,,\\
 C_\pm&=&\sin[\nu L\theta(\pm\delta)]\,,\\
\lambda_\pm &=&\frac{\hbar \sqrt{v_F^\pm}}{iE}\frac{\Delta}{\bar v}\frac {\nu}{\theta(\pm\delta)}\,.
\end{eqnarray}
The fraction of reflected flux is given by $\vert r\vert ^2$  and the transmitted flux
is $1-\vert r\vert ^2 = v^+_F\vert t\vert ^2/v^-_F$. The explicit evaluation 
(which is some what lengthy)
of the coefficients gives 

\begin{eqnarray}
 \vert r\vert^2&=&0\,,\\
\vert t \vert^2&=&\frac{v^-_F}{v^+_F}\,,
\end{eqnarray}
and therefore the electrons are totally transmitted across the heterojunction. Of course that this
result could have been anticipated from the conclusions of Sec. \ref{step-like}, since it is
 always
possible to represent a well-behaved function by a sum of infinitesimal rectangles.
(It is however elegant to have an exact solution to a given problem.)

\subsection{The linear velocity profile with massive particles}

\begin{figure}[!hbp]
\centering
\includegraphics[width=7cm]{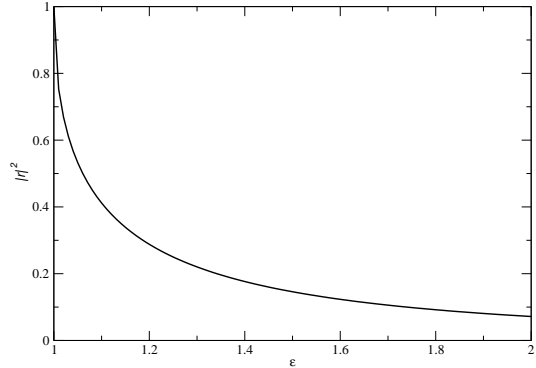}
\caption{Left: plot of Eq. (\ref{rwithmass}) as function of $\epsilon$.
Right: plot of Eq. (\ref{rwithmass}) as function of $\epsilon$ }
 \label{Fig_rwithmass}
\end{figure}

We can now use what we have just learned to a deal with a more general
situation where the electronic spectrum changes
due to a change of the material. The simplest case is that where for $x=\delta$ the mass
of the quasi-particles jumps from zero to a finite value. The connection between the two materials
is represented by the linear profile of the velocity discussed in Sec. \ref{linear}. The
mass profile is
\begin{equation}
m(x)=m\,\theta (x-\delta)\,,
\end{equation}
that is, the change is velocity takes place outside the region of finite mass.
Since the particle has a finite mass the spectrum changes to $E=\pm\sqrt{v^2_+q^2+m^2v^4_+}$,
and the wave function in this region changes to
\begin{equation}
\psi^{III}(x)=
t\left(
\begin{array}{c}
 \sqrt{E^2-m^2v^4_+}\\
 E-mv^2_+
\end{array}
\right)e^{-iqx}\,.
\label{psitransmassive}
\end{equation}
The overall modification is the replacement of $e^{iq}$ by $\sqrt{E^2-m^2v^4_+}e^{iq}$
in the third row, and of $e^{iq}$ by $(E-mv^2_+)e^{iq}$ in the fourth row.
Working out the calculation of $\vert r\vert^2$ we obtain

\begin{equation}
\vert r\vert^2= -[1 + 2 \epsilon (-\epsilon + \sqrt{\epsilon^2 - 1})]\,,
\label{rwithmass}
\end{equation}
with $\epsilon=E/(mv^2_+)$. As expected, the result does not depend on $\delta$ and
on $v^\pm_F$,
due to the reasons discussed in Sec. \ref{linear}. A plot of Eq. (\ref{rwithmass})
is given in Fig. \ref{Fig_rwithmass}. The situation is same if the mass profile had been chosen as
\begin{equation}
m(x)=m\,\theta (x+\delta)\,,
\end{equation}
this is, the linear profile rises up inside the massive region.

\subsection{The linear mass profile}
\label{linearmass}
We have seen above that the linear velocity profile does not contribute to the
backscattering of Dirac-fermions, even when it is combined with a region where
fermions are massive. In fact, any velocity profile produces no backscattering in one-dimension.
All the backscattering comes from the change in the dispersion,
due to the presence of the mass term. This fact motivates the question of what is form of the
wave function and of the coefficient $\vert r \vert^2$ if the mass does not change
abruptly but in a smooth way? A choice for the change of the mass profile, leading to an exact
solution, is that described by:

\begin{equation}
 m(x)= m\,\theta(x-\delta) + m\frac {x}{\delta} \theta(x)\theta(\delta-x)\,.
\label{masslinearprofile}
\end{equation}

We approach the solution of the scattering problem by solving first the Dirac equation
subjected a mass profile $m(x)=m\frac {x}{\delta}$. The method of solution is inspired
in that used for the 3+1 Dirac equation \cite{Cimento,chinesePhys}.
In this case the Fermi velocity is constant, and therefore we have to solve Eq. (\ref{HD})
with the additional term $\sigma_z mv^2_F\frac {x}{\delta}\psi$:
\begin{equation}
v_F\sigma_x\frac{\hbar}{i}\frac{d\psi}{d\,x}+\sigma_z mv^2_F\frac {x}{\delta}\psi=E\psi\,.
\label{Hlinearmass}
\end{equation}
The solution of this problem proceeds in several steps.  The first is to operate on the left
of Eq. (\ref{Hlinearmass}) with the operator
\begin{equation}
\frac{d}{d\,x}\sigma_y\sigma_z\,.
\end{equation}
After some algebra we obtain the following result
\begin{equation}
\hbar^2 v^2_F \frac{d\,^2\psi}{d\,x^2}=
-[(E^2-m^2v^4_F x^2/\delta^2)+\sigma_y\hbar m v_F^3/\delta\,]\psi\,.
\label{H2ndorder}
\end{equation}
The two components of the spinor are still coupled in Eq. (\ref{H2ndorder}).
In order to decoupled them we use the unitary transformation
\begin{eqnarray}
 \phi&=&U^\dag\psi\,,\\
U&=&
\left(
\begin{array}{cc}
 1 & 0\\
 0 & i
\end{array}
\right)\,.
\end{eqnarray}
The above transformation changes $\sigma_y$ to $\tilde\sigma_y=U^\dag\sigma_y U$,
which, still mixing the spinors in the Eq. (\ref{H2ndordertrans}) below, does it with the same sign 
(this is a crucial step). After applying the unitary transformation
we obtain
\begin{equation}
\hbar^2 v^2_F \frac{d\,^2\phi}{d\,x^2}=
-[(E^2-m^2v^4_F x^2/\delta^2)+\tilde\sigma_y\hbar m v_F^3/\delta\,]\phi\,.
\label{H2ndordertrans}
\end{equation}
We now introduce two new functions defined by $F_\pm=\phi_1\pm\phi_2$,
where $\phi_i$ (with $i=1,2$) are the components of  the spinor $\phi$.
In terms of the functions $F_\pm$ the eigenproblem takes
the form:
\begin{equation}
-\hbar^2 v^2_F \frac{d\,^2 F_\pm}{d\,x^2}+[m^2v^4_F x^2/\delta^2-\epsilon_\pm^2]F_\pm=0\,,
\label{H2F}
\end{equation}
where $\epsilon_\pm^2=E^2\pm \hbar m v_F^3/\delta$. The general solution of Eq. (\ref{H2F})
is given in terms of parabolic cylinder functions
 $D_\nu(x)$  (see  \ref{Ap0})

\begin{equation}
F_\pm(x) = A_\pm D_{\nu_1^\pm}(\sqrt 2(b/a)^{1/4}x)+
B_\pm D_{\nu_2^\pm}(i\sqrt 2(b/a)^{1/4}x)\,,
\label{generalsol}
\end{equation}
with 
\begin{eqnarray}
\nu_1^\pm&=&-\frac 1 2 + \frac{\epsilon_\pm^2}{2\sqrt{ab}}=\nu-(1\mp1)/2\,, \\
\nu_2^\pm&=&-\frac 1 2 - \frac{\epsilon_\pm^2}{2\sqrt{ab}}=-\nu_1^\pm-1\,\\
\nu &=& \frac{E^2\delta}{2\hbar v_F^3m}\,,\\
a &=& \hbar^2v_F^2\,,\\
b &=& v^4_Fm^2/\delta^2\,. 
\end{eqnarray}
If we were looking for the solution of the problem valid to all values of $x$,
the functions $D_\nu(x)$ with imaginary argument, since are not real 
and are not normalizable in the infinite volume, had to be excluded.
Therefore, in the limit $x\rightarrow \pm\infty$ the normalizable solutions
are those where $F_\pm(x)$
represents the one-dimensional harmonic oscillator wave functions (see
\ref{ApI}). 
Thefore the solution in the interval $x \in [-\infty,\infty]$ is
\begin{equation}
F_\pm(x) = 2A_\pm D_{\nu_1^\pm}(\sqrt 2(b/a)^{1/4}x)\,.
\end{equation}
This choice guaranties that the have function is well behaved at $x\rightarrow \pm\infty$
when $\nu_1^\pm$ is equal to a positiver integer (see \ref{ApI}).
Therefore, in the infinite volume, the solution of Eq. (\ref{H2ndorder}) takes the form
\begin{equation}
\psi(x)=
\left(
\begin{array}{c}
 A_+D_{\nu_1^+} + A_-D_{\nu_1^-}\\
 iA_+D_{\nu_1^+} -i A_-D_{\nu_1^-}
\end{array}
\right)\,, 
\label{secondDsol}
\end{equation}
where the argument of the  parabolic cylinder functions has been omitted for simplicity of writing.
We still have to check whether Eq. (\ref{secondDsol}) is solution of Eq. (\ref{Hlinearmass}).
Introducing the solution (\ref{secondDsol}) in Eq. (\ref{Hlinearmass}) we obtain that 
the wave function is a solution if
\begin{eqnarray}
\label{Aminus0}
A_-&=&\frac {\Lambda \nu}{E}A_+\,,\\
 \Lambda &=& v_F\hbar \sqrt{\frac{2v_F^2 m}{v_F\hbar\delta}}=v_F\hbar /\beta\,.
\label{Aminus}
\end{eqnarray}

\begin{figure}[!hbp]
\centering
\includegraphics[width=7cm]{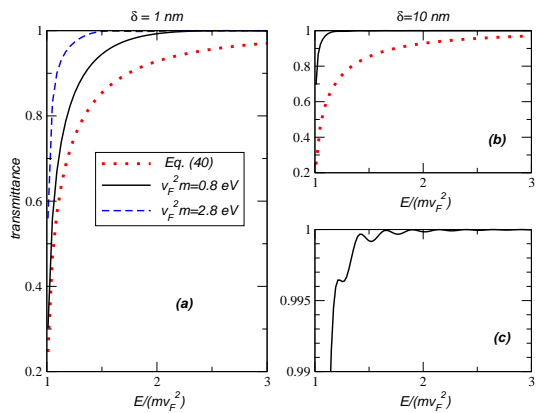}
\caption{Transmittance of Dirac electrons across a region where the mass
profile changes according to Eq. (\ref{masslinearprofile}). The dotted line
represents the transmittance in the case of an abrupt change of the mass profile,
as given by Eq. (\ref{rwithmass}). The energy gap 
is represented by the value of $mv^2_F$. In panel {\bf (a)} we give the results
for a value of $\delta$ of 1 nm. In panels {\bf (b)} and {\bf (c)} we plot the same
but for $\delta=$ 10 nm. In panel {\bf (c)} we give a zoom in of the solid curve
of panel {\bf (b)}; it is clear the existence of resonances in the transmittance.
The Fermi velocity is $10^6$ m/s, leading to $v_F\hbar=0.66$ eV$\cdot$nm.}
 \label{Fig_mass_sctt}
\end{figure}

What we have discuss so far assumes that Eq. (\ref{Hlinearmass}) holds for every
$x$. Our interest, however, is on the scattering problem of electrons when 
Eq. (\ref{Hlinearmass}) holds for $x\in]0,\delta[$. In this case, the general solution
(\ref{generalsol}) with both real an complex wave functions must be used.
The strategy is the same used before. We known at the outset that the solution of
Eq. (\ref{Hlinearmass}) is obtained from the solution (\ref{secondDsol}) by fixing
the value of a constant, as in Eq. (\ref{Aminus0}). The final solution is 

\begin{eqnarray}
 \psi(x)
&=& A
\left(
\begin{array}{c}
 D_\nu(x/\beta) +\nu\Lambda/E D_{\nu-1}(x/\beta)\\
iD_\nu(x/\beta)-i\nu\Lambda/E D_{\nu-1}(x/\beta)
\end{array}
\right)
\nonumber\\
&+&
B
\left(
\begin{array}{c}
 D_{-\nu}(ix/\beta) +i\nu\Lambda/E D_{-\nu-1}(ix/\beta)\\
-iD_{-\nu}(ix/\beta)-\nu\Lambda/E D_{-\nu-1}(ix/\beta)
\end{array}
\right)
\end{eqnarray}
which is valid in the region $x\in]0,\delta[$. As in Sec. \ref{linear},
we wave the incoming wave function given by Eq. (\ref{psiinc}), for 
$x<0$, and the transmitted one given by Eq. (\ref{psitransmassive}), for $x>\delta$.
The reflected and transmitted amplitudes $r$ and $t$, respectively, are obtained from imposing
the continuity of the wave function at $x=0$ and $x=\delta$.
In Figure \ref{Fig_mass_sctt} we give some numerical examples of our calculation.
It is clear from Fig. \ref{Fig_mass_sctt} that the smoothness of the change of the mass
profile leads to a larger transmittance, for a given energy, than when the mass-profile
changes abruptly. Also, when the change in the value of the mass takes place over
a relatively large regions, the transmittance shows the presence of resonances, as
can be seen in panel {\bf (c)} of Fig.  \ref{Fig_mass_sctt}.

\section{Discussion and conclusions}

We have studied several scattering problems using a modified version of the Dirac Hamiltonion
which incorporates the possibility of a spatial dependent velocity and mass terms. 
Two exact solutions were given. 
We showed that in the case of a velocity profile
it is necessary the modification of the original Dirac Hamiltonian, in order to have
a Hermitian problem.
We have shown that Klein
tunneling is not suppressed by a change on the velocity profile, with a transmittance equal to
unity always. This was understood by studying the case of an abrupt change in the velocity profile
and also by solving exactly the case where the velocity changes linearly across a given region.

We have also studied the case where the mass term depends on position. 
For this situation, we solved the cases of an abrupt change of the mass value and of a linear
change of the mass profile. In both cases we see the presence of backscattering, with 
values of the transmittance smaller than unity. The smoother mass profile induces less amount
of backscattering.
 
It is interesting to consider next the case of Dirac electrons in a strip of finite width $W$,
a situation relevant for graphene strips, and
see if in this case a position dependent velocity profile does produce backscattering. This
will be considered in a follow up publication. 

\section*{Acknowledgments}

This work was supported by FCT under the grant PTDC/FIS/64404/2006.
J.~M.~B. Lopes dos Santos and Vitor Pereira are  acknowledged for fruitful discussions.
J.~L. Martins is acknowledged for suggesting the form of the Hermitian
Dirac Hamiltonian. 

\appendix

\section{A derivation of the Dirac equation with a position dependent velocity }
\label{Ap00}
Consider a one-dimensional tight-binding Hamiltonian with two
atoms per unit cell such the the hopping parameter within a unit cell
$n$ is $-t_n$ and among nearest unit cells in $t_n$. The Hamiltonian
in second quantization reads
\begin{equation}
 H=\sum_n [-t_n(a^\dag_nb_n+b^\dag_na_n)+ t_n(a^\dag_nb_{n-1}+b^\dag_{n-1}a_n)]\,.
\label{eq:HTB}
\end{equation}
The in the case $t_n=t$ (the hopping is independent of the unit cell) the
spectrum of the electrons is 
\begin{equation}
 E_{\alpha=\pm}=\alpha 2t\vert \sin(kc/2)\vert\,,
\end{equation}
with $c$ is the length of the unit cell vector. Close to $k=0$ (zero energy) the spectrum is linear
in momentum and a massless Dirac spectrum is generated with a Fermi velocity given
by $v_F$. Let us now assume the general case of a $t_n$ dependent on the unit cell position
and obtain from the Hamiltonian (\ref{eq:HTB}) the effective field-theoretical model
that describes the system at low energy ($E\simeq 0$). Since the momentum close to which
the Dirac spectrum develops is $k=0$, we can write immediately the effective
field theoretical model as

\begin{equation}
H=-\frac 1 c\int dy\, t(y)\left[
a^\dag(y)c\frac{d\,b(y)}{d\,y}
+c\frac{d\,b^\dag(y)}{d\,y}a(y)
\right] \,,
\label{eq:Hfield}
\end{equation}
where we have used the expansion $b^\dag(y-c)\simeq b^\dag(y)+c d[b^\dag(y)]/d\,y$.
Integrating by parts the second term of Eq. (\ref{eq:Hfield}) and using the Pauli
matrices to help condensing the results we obtain
\begin{equation}
H=\frac{1}{c}\int dy \left[
\Psi^\dag\sqrt{v_y(y)}\sigma_yp_y\sqrt{v_y(y)}\Psi + 
\Psi^\dag\frac{c\sigma_x}{2}\frac{d\,t(y)}{d\,y}\Psi
\right] \,
\label{eq:HfieldII}
\end{equation}
with $v(y)=ct(y)/\hbar$ and $\Psi^\dag=[a^\dag(y) b^\dag(y)]$. The
first term in Eq. (\ref{eq:HfieldII}) is our proposed Hamiltonian
(\ref{HDII}).

\section{Weber's differential equation}
\label{Ap0}
We give here some basic information  on the Weber's differential equation ,
aiming to give the text
a selfcontain nature and to fix notation and definitions.
Weber's differential equation is defined as
\begin{equation}
 y''(z)+(\nu + 1/2-z^2/4)y(z)=0\,,
\label{weber}
\end{equation}
and its two independent solutions are the  parabolic cylinder functions $y(z)=D_\nu(z)$
and $y(z)=D_{-\nu-1}(iz)$. Equation (\ref{H2F}) is of the general form
\begin{equation}
y''(x)+(-ax^2+c)y(x)=0\,.
\label{ours}
\end{equation}
Making the transformation $x=\beta z$, with $\beta$ given by
$\beta=(4a)^{-1/4}$ and $\nu=-1/2+c\beta^2$, we reduce Eq. (\ref{ours})
to Weber's equation (\ref{weber}). Using for $a$ and $c$ the particular
values of our problem we obtain
\begin{equation}
 \beta = \sqrt{\frac{\hbar v_F\delta}{2v^2_Fm}}\,,
\end{equation}
and
\begin{equation}
 \nu = -\frac 1 2 + \frac{\epsilon^2_\pm\delta}{2\hbar v_F^3m}\,.
\end{equation}
The derivative of the  parabolic cylinder functions obeys
\begin{eqnarray}
D'_\nu(z)+zD_\nu(z)/2-\nu D_{\nu-1}(z)&=&0\,. \\
D_{\nu+1}(z)-zD_\nu(z)+\nu D_{\nu-1}(z)&=&0\,.
\end{eqnarray}
Using the results of Ref. [\cite{abramowitz}],
the solution of $D_{\nu}(z)$ can be written in terms of the Kummer
confluent hypergeometric function, $U(a,b,x)$, as
\begin{equation}
D_\nu(z)=2^{\nu/2}e^{-z^2/4}U(-\nu/2,1/2,z^2/2)\,. 
\end{equation}
The Kummer function $U(a,1/2,z)$ is computed using the 
the Kummer
confluent hypergeometric function, $M(a,b,x)$, as

\begin{equation}
 U(a,1/2,z)= \sqrt{\pi}\,\frac{M(a,1/2,z)}{\Gamma(a+1/2)}-2\sqrt{z\pi}\,
\frac{M(a+1/2,3/2,z)}{\Gamma(a)}
\,.
\end{equation}

\section{Eigenvalues of the scalar potential $V(x)=\sigma_z m v^2_F x/\delta$}
\label{ApI}

The problem we introduced in Sec. \ref{linearmass} was that of a particle that moves
in a heterosctructure with a mass dependent position. We can, however, think of this problem
as that of a Schr\"odinger particle moving in the scalar potential $V(x)=m^2 v^4_F(x/\delta)^2$.
Since Eq. (\ref{H2F}) is that of an one-dimensional harmonic oscillator, 
the normalizable solutions have the well known form 
\begin{equation}
z_n(u)=\frac{\pi^{-1/4}}{\sqrt{2^n n!}}e^{-u^2/2} H_n(u)\,,
\end{equation}
where 
\begin{equation}
 u=x\left(
\frac {V_F^2m}{\delta V_F\hbar}\,,
\right)^{1/2}
\end{equation}
upon the identification
\begin{eqnarray}
 \frac {1}{m_0}&\leftrightarrow& 2v^2_F\,,\\
\omega_0^2&\leftrightarrow& 4v^6_Fm^2/\delta^2\,,
\end{eqnarray}
with $m_0$ and $\omega_0$ the mass and the frequency of the oscillator, respectively.
The spectrum is obtained from $\epsilon^2_\pm =2\hbar v^3_Fm(n+1/2)/\delta$,
with $n=0,1,2,\ldots$, leading to
\begin{eqnarray}
 E^2_+&=&2\hbar v^3_Fm n\delta^{-1}\,,\\
 E^2_-&=&2\hbar v^3_Fm (n+1)\delta^{-1}\,.
\end{eqnarray}
Since the energy has to be same for both $F_+$ and $F_-$,
we choose the solutions \footnote{It is interesting to note that the system
supports a zero energy mode:$F^{n=0}_+=z_0$ and $F^{-1}_-=0$. }
$F^n_+=z_n$ and $F^n_-=z_{n-1}$,
with $z_{-1}=0$. 
Finally, from the definition $F_\pm=\phi_1\pm\phi_2$,
we obtain (properly normalized)
\begin{eqnarray}
\phi_1(x)=[z_n(x)+z_{n-1}(x)]/\sqrt 2\,,\\
\phi_2(x)=[z_n(x)-z_{n-1}(x)]/\sqrt 2\,.
\end{eqnarray}
The above equations are the solution of Eq. (\ref{H2ndorder}).

\section*{References}


\end{document}